\documentclass[final,10p,times]{elsarticle}
\usepackage[letterpaper, portrait, margin=0.8in]{geometry}
\usepackage{graphics}
\usepackage{graphicx}
\usepackage{epsfig}
\usepackage{amssymb}
\usepackage{amsthm}
\usepackage{url}
\usepackage{caption}
\usepackage{subcaption}
\usepackage{caption}
\biboptions{numbers,sort}
\begin{document}

\begin{frontmatter}

\title{ReHabgame: A non-immersive virtual reality rehabilitation system with applications in neuroscience}

\author[mymainaddress]{Shabnam Sadeghi Esfahlani\corref{mycorrespondingauthor}
\cortext[mycorrespondingauthor]{Corresponding author}}
\ead{shabnam.sadeghi-esfahlani@anglia.ac.uk}
\author[mymainaddress]{Tommy Thompson}\ead{tommy@t2thompson.com}

\author[mysecondaddress]{Ali Davod Parsa}\ead{ali.parsa@anglia.ac.uk}
\author[mymainaddress]{Ian Brown}\ead{ian.brown@anglia.ac.uk}
\author[mymainaddress]{Silvia Cirstea}\ead{silvia.cirstea@anglia.ac.uk}
\address[mymainaddress]{Anglia Ruskin University, Department of Computing and Technology, CB5 8DZ, Cambridge, United Kingdom}
\address[mysecondaddress]{Anglia Ruskin University, Department of Medical Science, CB5 8DZ, Cambridge, United Kingdom}

\begin{abstract}
This paper proposes the use of a non-immersive virtual reality rehabilitation system "ReHabgame" developed using Microsoft Kinect$^{TM}$ and the Thalmic$^{TM}$ Labs Myo gesture control armband. The ReHabgame was developed based on two third-person video games that provide a feasible possibility of assessing postural control and functional reach tests. It accurately quantifies specific postural control mechanisms including timed standing balance, functional reach tests using real-time anatomical landmark orientation, joint velocity, and acceleration while end trajectories were calculated using an inverse kinematics algorithm. The game was designed to help patients with neurological impairment to be subjected to physiotherapy activity and practice postures of daily activities. The subjective experience of the ReHabgame was studied through the development of an Engagement Questionnaire (EQ) for qualitative, quantitative and Rasch model.\\
The Monte-Carlo Tree Search (MCTS) and Random object (ROG) generator algorithms were used to adapt the physical and gameplay intensity in the ReHabgame based on the Motor Assessment Scale (MAS) and Hierarchical Scoring System (HSS). Rasch analysis was conducted to assess the psychometric characteristics of the ReHabgame and to identify if these are any misfitting items in the game.\\
Rasch rating scale model (RSM) was used to assess the engagement of players in the ReHabgame and evaluate the effectiveness and attractiveness of the game. The results showed that the scales assessing the rehabilitation process met Rasch expectations of reliability, and unidimensionality. Infit and outfit mean squares values are in the range of $(0.68-1.52)$ for all considered 16 items. The Root Mean Square Residual (RMSR) and the person separation reliability were acceptable. The item/person map showed that the persons and items were clustered symmetrically.
\end{abstract}


\end{frontmatter}

\section{Introduction}
Rehabilitation is a complicated and active process by which those affected by injury or disease achieve a full recovery or, if a full recovery is not possible, realise their optimal physical, mental and social potential that allows them to regain much of their independence and quality of life \cite{ortiz2014virtual}. It is agreed that intensive, repetitive, and goal-oriented rehabilitation process improves motor function and cortical reorganisation in stroke patients with both acute and long-term (chronic) impairments. However, this recovery process is typically slow and labour-intensive, usually involving extensive interaction between one or more therapists and one patient \cite{kan2011development}. Motor disability does not only cause limitations in functional motor control, muscle strength, and range of motion (ROM) but can limit the ability to perform daily tasks, that could cause isolation, and reduce participation in community activities \cite{nowak2008impact} and \cite{piron2009assessment}. Participating in repetitive exercises can help these people to overcome the limitations they experience, but lack of action and isolation stops them to perform recommended practices thus, become weaker, and cause obesity-related chronic health conditions and so on. Presumably, every person participating in rehabilitation hopes to be happily situated, productively occupied, and efficiently supported throughout the process \cite{mccoll2001community}. A lack of motivation is another impediment that stops them to participate in physiotherapy sessions regularly. \\
Studies in computational neuroscience have shown that virtual reality (VR) rehabilitation techniques could create an environment that recovers health conditions as well as community integration \cite{turolla2013virtual} and \cite{piron2009assessment}. A study by \cite{turolla2013virtual} among post-stroke patients was conducted to investigate the effects of the two treatment methods: (I) combined VR and conventional therapy and (II) standard therapy alone. Their findings showed that the first process (combined VR and traditional treatment) was more efficient than the other. The past decade has witnessed the creation of a vast number of successful VR based rehabilitation, educational and commercial video games. They use new technology to simulate realistic features of the real world, \cite{mohanty2011teaching}, \cite{weiss2013video}, \cite{mirelman2009effects}, \cite{o2013cognitive}, \cite{ustinova2014virtual}, \cite{levin2015emergence}, \cite{saposnik2010effectiveness}, \cite{dobkin2004strategies}, \cite{tsekleves2016development}, \cite{smith1999task}, \cite{page2004efficacy} and \cite{holden2002virtual}. \cite{piron2001virtual} conducted the application of VR in psychology as an assessment tool considered as a highly sophisticated form of adaptive testing. \cite{holden2005virtual} reviewed the applications of VR in the field of motor rehabilitation. They reported that the people with disabilities appear capable of motor learning within VR environments and the movements learned in VR are transferable to real-world equivalent. \cite{mirelman2009effects} investigated the effects of training of the lower extremity using a robot coupled with virtual reality (VR). They also showed that movements learned in a VR environment are transferable to real-world similar motor tasks. \cite{villiger2011virtual} have used interesting VR scenarios to reshape cortical networks in patients which resulted in significant improvement in controlling the body, a decrease in neuropathic pain, and an increase in muscle strength. 
\begin{figure*}[thpb]
     \centering
     \includegraphics[scale=0.8]{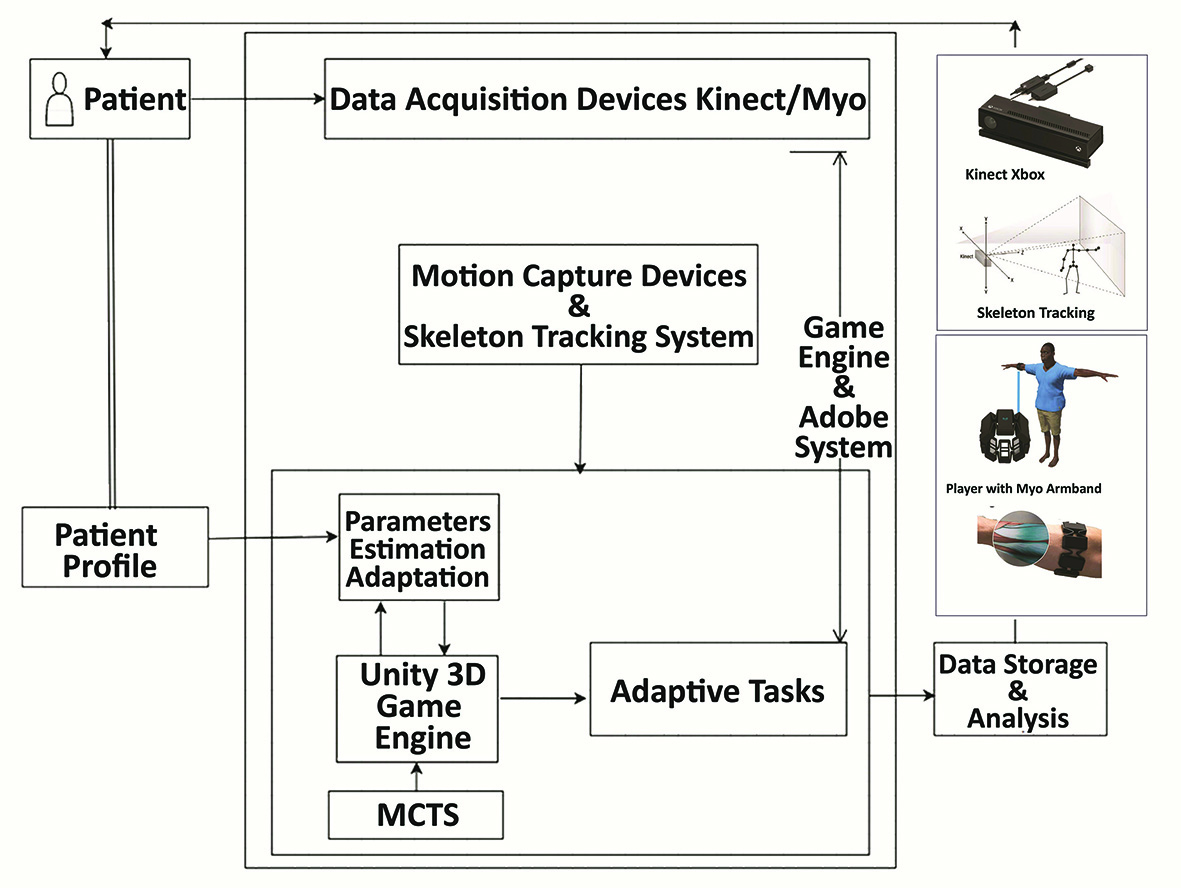}
      \caption{The ReHabgame diagram and its components with MCTS algorithm.}
      \label{architecture}
   \end{figure*}
The "ReHabgame" was developed based on the neuromuscular rehabilitation system that assists players to combat the physical impairment and improve the upper limb performance. It is a flexible, non-immersive, low-cost VR system that does not use any head mounted devices but simulates the VR features integrating off-the-shelf devices such as the Kinect V2 (Microsoft in 2014)\footnote{https://en.wikipedia.org/wiki/Kinect} and Myo armband (Thalmic Labs in 2016). This system encourages motor impaired participants to interact with virtual objects which are spawned in the scene. This system provides an easy and engaging rehabilitation process compared to using other haptic devices. It could be used by patients for long-term recovery which leads to better rehabilitation outcome. It was developed in Unity game engine that integrates the motion/gesture capture, image processing, skeleton tracking, IR emitter, inertial measurement units and EMG sensors capabilities into the games. The postural control/balance, gestures and range of motion (ROM) of the upper body was tracked through this marker less system. It facilitates the calibration of joints orientation, position of the leading joints and muscle activities throughout the gameplay. The ReHabgame was built as a single-player (the clinician's avatar instructs the patient's avatar to copy the exercises) and two-player game (two patients play at the same time and compete against one another). All the data were collected and saved in the background (hard drive) through the devices and were used for qualitative and quantitative analysis.\\
In this study, we define VR as a computer-based, interactive, multi-sensory environment that occurs in real-time. It provides enhanced feedback about movement characteristics, and advanced motor task learning and execution. Our study shows that the use of interactive VR system improves upper limb rehabilitation and habilitation which concurs with findings in literature including \cite{piron2001virtual, piron2005virtual, holden2005virtual, piron2009exercises, fluet2010interfacing}.
Fig~\ref{architecture} illustrates the architecture of the system with its components. Two algorithms were adapted to acquire the gameplay to players’ ability using MCTS algorithm and Random Object Generator (ROG). Players interact with the virtual objects in the virtual park through three scenarios (case studies) such as; (I) "Reach-Grasp-Release fruits game", (II) "Reach-Press-Hold buttons game" and (III) "Reach-Press buttons game" which are analysed together as the ReHabgame. An Engagement Questionnaire (EQ) was developed and administered to a group of 20 subjects to assess subjectively the effectiveness and efficiency of the ReHabgame using item response theory (IRT). The results of the Rasch analysis of the outcomes of this experiment are presented in this paper. To our knowledge, this is the first study that proposes an instrument to measure the subjective experience and employs Rasch analysis to evaluate the engagement in a motor rehabilitation game. Three features of game engagement were considered: flow, presence and absorption \cite{brockmyer2009development, ryan2006motivational}. The Rasch model analysis of the gameplay shows that there is an appropriate balance between skill and challenge (flow). Participants feel present in the game and perceive a high-level of control (presence), which in turn offers a raised degree of engagement, thus, distracting them from the pain associated with the impairment (absorption).
\section{Design}
\subsection{ReHabgame principles, design and operation}
The player reaches and grasps virtual fruits generated in the 3D world and releases them above the virtual basket \cite{esfahlani2017adaptive, esfahlani2016intelligent}. The system is designed for one or two players to play at the same time independently as illustrated in Fig~\ref{Fig.TwoPlayer}. It can be set up to rehabilitate the upper limbs (right hand, the left hand or both hands at the same time) and postural balance. The system combines the abduction, adduction, flexion, extension, internal/external rotation and horizontal abduction/adduction. The two-player game may be arranged as a clinician (first avatar/ instructor) shows the movements of the patient (second avatar) in that the patient follows the actions of the clinician and accordingly the patient's ROM is recorded by the system. These data are used by the artificial intelligence (AI) control system to generate virtual objects within patient's comfort zone initially and then gradually push them further out of the comfort zone. The algorithm is designed based on inverse kinematic in that the joints (such as head, neck, shoulder, elbow, wrist, etc.) location are monitored through the Kinect for further analysis. The ReHabgame was developed as follows:
\begin{enumerate}
\item \textbf{"Reach-Grasp-Release"}: The players gather the fruits spawned in the scene and release them above the virtual baskets to gain points. As soon as the fruit is collected it must be released above the basket to hit the bottom of the basket to collect points. It was developed using the gravity component and raycast hit function in Unity. Fig~\ref{Fig.TwoPlayer-a} shows the female avatar who has collected the orange with the left hand and aims to release it above the basket. The downward red circles indicate that it shouldn't be released unless the downward circles changed to green colour. Thus, the player should adjust her arm to receive positive visual feedback (the colour changes from red circles to green). The male character's shoulder joint follows a horizontal abduction (extension), the elbow joint is in a flexion mode and the hand is extended. The male avatar's head and neck are tilted towards right shoulder that is detected as a problem in the postural control system. The streamed video and joint locations collected via the Kinect are used for further analysis. Fig~\ref{Fig.TwoPlayer-b} illustrates the female avatar who aims to reach the kiwi while the right shoulder's joint is abducted. The male character in this figure has picked the peach and wants to release it above the basket but the visual feedback indicating the red downward circles which means it is not safe to release it. In Fig~\ref{Fig.TwoPlayer-c} the female avatar collected the strawberry but cannot release unless the circles indicating green. The male character is idle in this figure. Fig~\ref{Fig.TwoPlayer-d} shows the female avatar has collected the strawberry and wants to release it since the visual feedback specifies the green circles meaning that she can release the fruit to gain points. The male's avatar hand is extended and the elbow join is flexed.
\item \textbf{"Reach-Press-Hold"}: In the reach-and-press button game, the player reaches out to a button to press it; the latter task can be configured to require a press-and-hold action or a press-for-specific-time action from the player in that the steadiness and the vibration of the hand are recorded based on $30$ fps. \\
Fig~\ref{Fig.Button} illustrates the player \& avatar in button Reach-Press-Hold game. Buttons are generated in a matrix format defined in the system based on patient's range of movements (ROM). Fig~\ref{Fig.Button-a} shows the button generated around the player's head height in that the arm is abducted (away from the midline of the body). The player should raise the hand above the shoulder's height to reach the button which requires higher ROM.  Fig~\ref{Fig.Button-b} illustrates the button generated in an area around the player's heap height with arm abduction. Fig~\ref{Fig.Button-c} shows the arm's flexion (horizontal adduction) to reach the button generated at the shoulder's height. The player reached the button and pressed it for one second with a visual feedback via the green dot around the white circle. Fig~\ref{Fig.Button-d} illustrates the red part of the button is pressed for three seconds with three green visual feedbacks appeared around the button. If the player's press and hold is not steady enough, the timer restarts itself.

\begin{figure}
\centering
    \begin{subfigure}{0.41\textwidth}
        \centering
        \includegraphics[width=\linewidth]{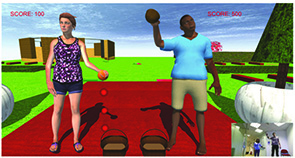}
        \caption{}\label{Fig.TwoPlayer-a}
    \end{subfigure} %
    \begin{subfigure}{.4\textwidth}
        \centering
        \includegraphics[width=\linewidth]{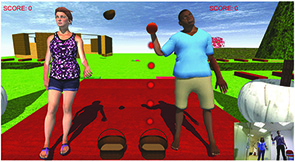}
        \caption{}\label{Fig.TwoPlayer-b}
    \end{subfigure} %
       \begin{subfigure}{.4\textwidth}
        \centering
        \includegraphics[width=\linewidth]{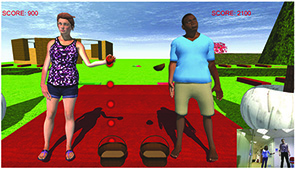}
        \caption{}\label{Fig.TwoPlayer-c}
    \end{subfigure}
       \begin{subfigure}{.41\textwidth}
        \centering
        \includegraphics[width=\linewidth]{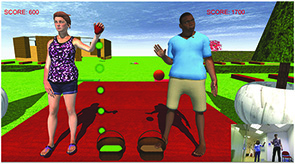}
        \caption{}\label{Fig.TwoPlayer-d}
    \end{subfigure}
\caption{Screen shots of a male and a female players with the representative avatars in the “Reach-Grasp-Release” fruits game. (a) shows the female avatar collected the peach and aiming to release it above the basket, and the male avatar collected a kiwi, (b) displays the female avatar aiming to collect the kiwi and the male avatar has collected the peach, (c) the female has collected the fruit and the male avatar is idle and (d) demonstrations the female avatar released the strawberry above the basket with green dots provided the positive feedback.}
\label{Fig.TwoPlayer}
\end{figure}

\begin{figure}
\centering
    \begin{subfigure}{0.4\textwidth}
        \centering
        \includegraphics[width=\linewidth]{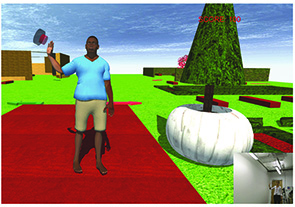}
        \caption{}\label{Fig.Button-a}
    \end{subfigure} %
    \begin{subfigure}{.39\textwidth}
        \centering
        \includegraphics[width=\linewidth]{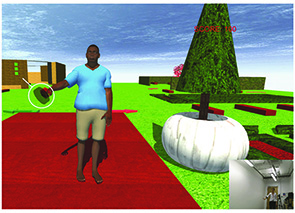}
        \caption{}\label{Fig.Button-b}
    \end{subfigure} %
       \begin{subfigure}{.4\textwidth}
        \centering
        \includegraphics[width=\linewidth]{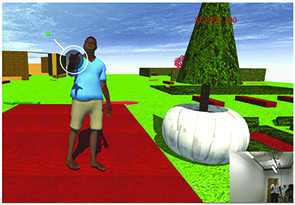}
        \caption{}\label{Fig.Button-c}
    \end{subfigure}
       \begin{subfigure}{.395\textwidth}
        \centering
        \includegraphics[width=\linewidth]{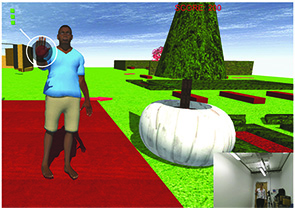}
        \caption{}\label{Fig.Button-d}
    \end{subfigure}
\caption{Screen shots of a player in the “Reach-Press-Hold” button game. (a) the avatar's right hand is engaged with the game which aiming towards the button to press its red part, (b) illustrates the avatar pressed the red button for three sec with three green dots appeared next to it which illustrates the visual feedback, (c) displays the avatar reached and pressed another button and holds it for one sec and (d) demonstrates the avatar with three seconds holding with its visual feedback.}
\label{Fig.Button}
\end{figure}
\end{enumerate}  
\subsection{Virtual object generator algorithms}
Monte Carlo tree search algorithm (MCTS) has revolutionised the game's performance where partial information requires decision making under uncertainties \cite{chaslot2008monte}. MCTS is used as a method of finding optimal choices in a given domain by taking random samples in the decision space and building a search tree accordingly. The primary mechanism of generating adaptive patient-specific tasks in the ReHabgame resides in the object generator algorithm, which employs the MCTS procedure and Random Object Generator (ROG). To explore the effects of physical and gameplay intensity the MCTS object generation algorithm was used versus ROG. \\
MCTS was based on Monte Carlo sampling with a large number of random simulated games. MCTS makes the computation converge to the right value much more quickly than simple Monte Carlo itself~\cite{ciancarini2010monte}. This probabilistic algorithm uses lightweight random simulations of a problem space to grow a search tree that represents the unique states of the problem domain, with child nodes representing the outcome of specific actions~\cite{browne2012survey}. In the ReHabgame the algorithm was broken down into four stages as highlighted below:
\begin{enumerate}
\item \textbf{Selection:} The algorithm starts at the root node, builds a tree of possible future game states. Each node in the tree represents a state and each link represents an action taken in that state and leads to a new state. While the state is built in the tree, the next action is chosen according to the statistics stored, in a way that balances between exploitation and exploration. The task is either selecting an action that leads to the best results so far called exploitation or a less likely action still needs to explore (known as exploration due to the uncertainty of the evaluation).
\item \textbf{Expansion:} More state (child nodes) needs to be added as new nodes to expand the tree. The tree is grown by one node for each simulated game according to the available actions. 
The first action is chosen by an expansion strategy and subsequently simulated. This results in a new game state, for which a new node is created. 
\item \textbf{Simulation:} After expansion, a rollout is done from the new node, which means that a simulation is run from the new node applying random actions until a pre-defined stop criterion is met or the game ends. The satisfactory weighting of action selection probabilities has a significant effect on the level of play. If all legal actions are selected with equal probability, then the strategy played is often weak, and the level of the Monte-Carlo program is sub-optimal \cite{chaslot2008monte}.
\item \textbf{Backpropagation (backup):} The score difference resulting from the rollout is backed up to the root node, which means that the reward is saved to the visited nodes then a new iteration starts. Backpropagation updates node statistics that are used for future decisions and ultimately inform the final decision made by the system for the root node.
\end{enumerate}
An effective variant and selection strategy is the modified version of MCTS called upper confidence bound for trees (UCT) controller that construct statistical confidence intervals, \cite{de2016monte, frydenberg2015investigating}. UCB is efficient and guaranteed to be within the best possible bound to address the exploration-exploitation in MCTS \cite{browne2012survey} as illustrated in Eqn. 1. A child node $j$ is selected to maximise UCT, where $n$ is the number of times the current node (parent) has been visited, $n_j$ is the number of times child $j$ has been visited and $C_p>0$ is a constant that shifts priority from exploration to exploitation.
$$
UCT= \bar X_j+C_p \sqrt {\frac{\ln n}{n_j}}
\eqno{(1)}
$$
By increasing $C_p$, the priority is shifted to exploration which means; states that have been visited less will be visited with a higher priority than states that have been visited more. A decrease changes priority to exploitation means that; states which have a great value are visited more to maximise reward. If more than one child node has the same maximal value, the tie is usually broken randomly. The estimated value for every node that has been visited in the iteration is updated with the reward. The estimated value of a node is the average of all rewards backed up to that node.\\
\section{Model}
\subsection{Motor Assessment Scale}
\begin{figure}[thpb]
      \centering
     \includegraphics[scale=2.5
     ]{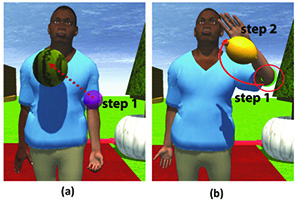}
      \caption{Avatar in “Reach-Grasp-Release” fruits game with two steps elbow and hand position; (a) shows step 1: place the elbow on purple button in that the colour changes to green as shown in (b), it also illustrates the step 2 in that the avatar moved the hand towards the virtual lemon's location.}
      \label{Fig.TwoStep}
   \end{figure}

The Motor Assessment Scale (MAS) \cite{sabari2005assessing, sabari2014rasch} is the standardised assessment tool in the clinical evaluation and outcome studies of stroke rehabilitation. It consists of 8 items representing different areas of motor behaviour: supine-to-side lying onto entire
side; supine to sitting over a side of the bed; balanced sitting; sitting to standing; walking; upper-arm function; hand movements; and advanced hand activities. The four items of MAS was used in this study for the design of VR system. The postural balance (increments of displacement of the person’s centre of mass and upper body joints) were monitored by the Kinect, upper-arm function; hand movements; and advanced hand activities data were assessed via both the Kinect and Myo armband devices.\\
A hierarchical scoring system (HSS) was used in ROG in that when the player achieves the criterion for a score of more difficult level the easier steps do not need to be tested. The study considers that the force of gravity and higher degrees of freedom of joints require greater control thus considered as advanced steps in the game level design \cite{sabari2005assessing}\\
\begin{itemize}
\item \boldmath{Upper-Arm Function}: Shoulder's depression, elevation, protraction, and retraction while placing elbow in a predefined position. Add elbow's Flexion and extension at different angles. Take hand to an area around a head position, and behind it. 
Fig~\ref{Fig.TwoStep}a shows the elbows’ fixed position that was selected by the virtual purple button, and Fig~\ref{Fig.TwoStep}b illustrates the hands’ orientation that was picked by the player. The fruit cannot be collected unless the purple button was pressed by the elbow (which was fixed to this location) and changed to green colour as an indicator of a positive feedback.\\
\item \boldmath{Hand Movements:} Radial deviation of the wrist, lift the hand, reach forward (extend arms fully/ partially) in a pickup shape. Show palm to the screen while shoulder and elbow are in different positions (i.e. shoulder protracted, elbow extended, wrist neutral or extended and palm in contact with virtual objects).\\
\item \boldmath{Advanced Hand Activities:} Grasp and releasing an object (I) the hand transport for carrying the hand toward the purpose considering shoulder, elbow and wrist movements and (II) the grasp component for shaping hand (fingers and thumb) for collection \cite{jeannerod1984timing} and \cite{nowak2008impact}. Hand gestures to pick up, and release open/close fingers and the thumb.
\end{itemize}

The scoring system was designed based on these components. If the attempt was successful the score is one: $+1$, if partially completed is half: $0.5$, if was unsuccessful is zero: $0.0$, as defined in Eqn. 2:
$$
 M_{Score}^i= 
 \left\{
 \begin{array}{lll}
M_{successful}^i & \ Score \ is \ +1.0\\
M_{Partially Successful}^i & Score \ is \ +0.5\\
M_{Not Successful}^i & \ Score \ is \ ~0.0\\
              \end{array}
              \right.
              \eqno{(2)}
$$

$M_{Score}^i$ is the score of the $i^{th}$ play where the setting could be graded for implementing the time as another factor to measure the count. It was determined by confronting the player's time with a linear interpolation between a best possible time (max score) and a maximum period allowed (0 counts).\\

 \begin{figure*}[thpb]
      \centering
     \includegraphics[scale=1.0  
     ]{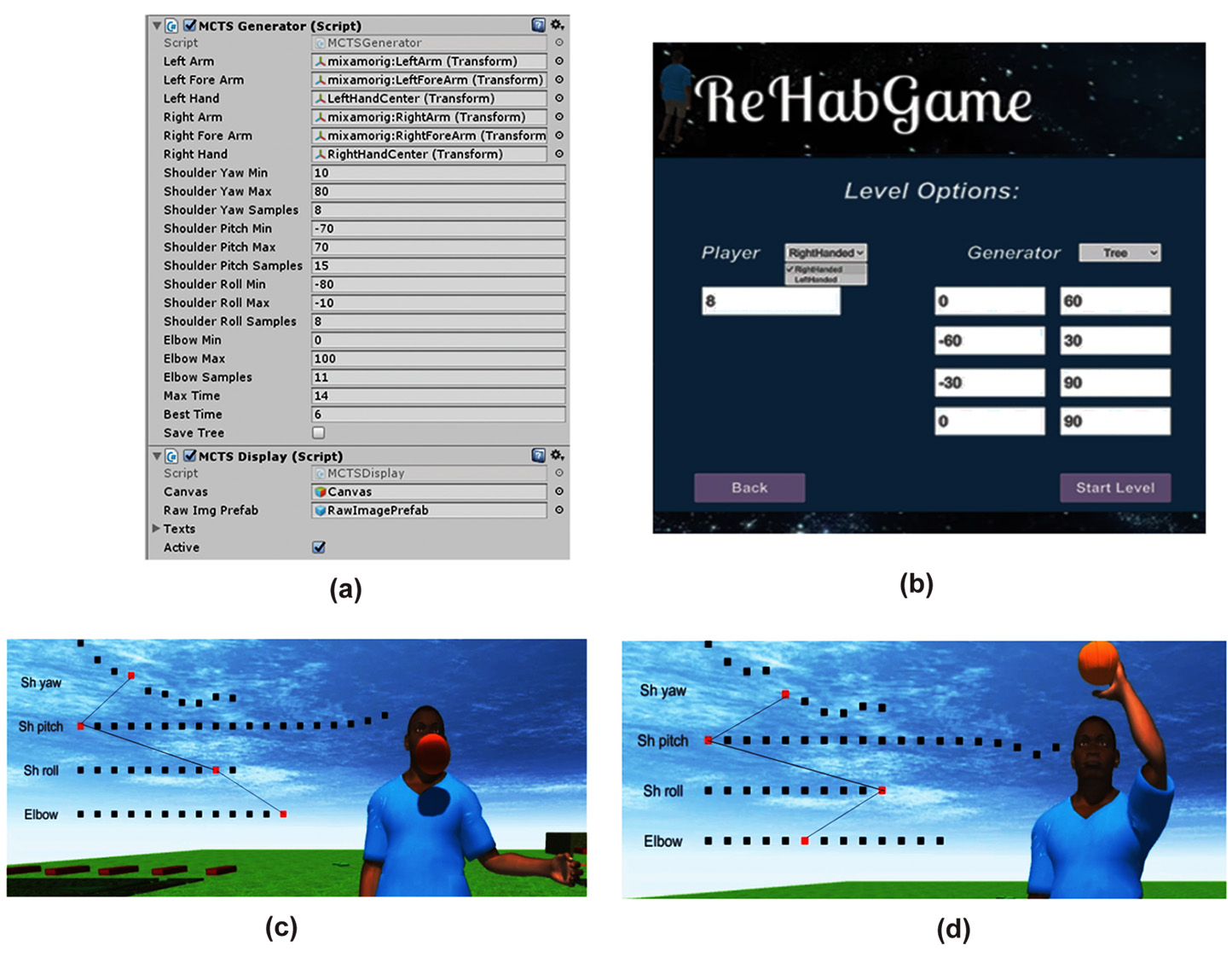}
      \caption{(a) is the game engine with the MCTS generator script that is attached to the generator object. (b) Illustrates the main menu of the ReHabgame. (c) shows the virtual peach generated in the player's neck high with the shoulder (sh-yaw, sh-pitch, sh-roll) and Elbow orientation values next to the player (these orientation are based on the red dots shown on the screen on avatar's right-hand side) and (d) illustrates the orange generated above player's head height.}
      \label{Fig.roll}
   \end{figure*}
The choice of the MCTS/ROG algorithms and other parameters including the number of iterations left/right-handed gameplay, type of the game and single/double player game can be defined through the main menu. Fig~\ref{Fig.roll}a and Fig~\ref{Fig.roll}b, illustrates the Game Engine and the main menu of the ReHabgame, respectively. The maximum and minimum range of orientation and movement are predefined (i.e. in the MCTS the shoulder's yaw is in the range $[0,90]$. The pitch is $[-90,90]$, the roll of the shoulder is between $[-90,0]$ and the elbow is within $[0,120]$. Each range is given a density based on the number of samples selected from the menu. This limited number of options allows for a non-infinite search-space. Fig~\ref{Fig.roll}c, Fig~\ref{Fig.roll}d shows the virtual fruit that is spawned in the 3D-world based on the confidence of the search-tree in the pre-established range of the upper limb orientation. The black and red dots, next to the avatar in Fig~\ref{Fig.roll}, show the effects of the parameters picked for shoulder and elbow freedom of movement. $10$ samples were selected for shoulder's yaw range (Sh yaw: first row of the black and red dots), $19$ samples for its pitch (Sh pitch: second row of the black and red dots), $10$ samples for its roll (Sh roll: third row of the black and red dots) and $13$ for the elbow (Elbow: fourth row of the black and red dots). The red dots in each picture corresponds to the orientation sample picked by the exploration tree to spawn the fruit. In Fig~\ref{Fig.roll}c the red dot in the "sh yaw" row represents the desired rotation of $38 ^{\circ}$ needed to be explored. Likewise, the "sh pitch", "sh roll" and "elbow" are determined based on their range, samples and confidence.\\
\subsection{Inverse Kinematics (IK)}
Kinematics refers to the mathematical description of motion without considering the underlying physical forces. The kinematics of the human body is specifically concerned with formulating and solving for the translation, rotation, position, and velocity of each body segment in real-world motions.
The movement of a kinematic chain, whether it is a robot or an animated character, is modelled by the kinematic equations of the chain \cite {papaleo2012inverse, tolani1996real, lura2012creation}. The IK determines the typical parameters that provide a desired position of the end-effector to achieve the task and is known as action planning \cite{lura2012creation}. IK techniques provide direct control over the placement of an end effector object at the end of a kinematic chain of joints, solving for the joint rotations which place the object at the desired location. IK is adapted for calculating the end effector for the virtual object's location. The algorithm that was used to calculate the joint angles required to collect the objects is defined in Eqn. 3:
$$
\begin{array}{llllll}
\theta_1\leftarrow \arctan2(y,x)\\
\theta_2\leftarrow \arctan2(z-l_1 , \sqrt[]{x^2+y^2})-\arctan2(l_3s_3 , l_2+l_3c_3) \\
\theta_3\leftarrow \arctan2(s_3,c_3)\\
\\
where:  \left\{
\begin{array}{ll}
c_3\leftarrow \frac{x^2+y^2+(z-l_1)^2-l_2^2-l_3^2}{2l_2l_3}\\\\
s_3\leftarrow  \sqrt[]{1-c_3^2}
\end{array}
\right.
\end{array} 
\eqno{(3)}
$$
Where $\theta _i$ are orientation around each axis and $l_i$ are each segment's length.
\section{Methodology}
\subsection{Design of the EQ}
Deep engagement in game-playing has the potential to be an important factor to predict the impact of using video games for rehabilitation purposes. Three key concepts are used in the literature to describe subjective experience relevant to engagement in a game: flow, presence and absorption \cite{brockmyer2009development, ryan2006motivational}. \\
\textbf{Flow} describes the enjoyment feelings due to a balance between skill and challenge in the process of performing an exercise \cite{csikszentmihalyi1992optimal, sinclair2009exergame}. The dual flow model measures a combination of attractiveness in terms of compelling gameplay and effectiveness in terms of physical outcomes \cite{sinclair2009exergame}. 
Defining a specific goal and an immediate performance feedback increase the likelihood of flow, and being in a flow state seems to enhance learning. Flow states also include a feeling of being in control, being one doing the activity, and experiencing time distortions.\\
\textbf{Presence} is another intrinsic motivation in gaming contexts which is the sense of being within the game world by controlling the game, a need for challenge and feeling of effectence (the feeling of efficacy \cite{white1959motivation}). Concepts of presence are widely discussed by game designers, who attempt to make the experience of virtual worlds feel real and authentic \cite{ryan2006motivational}.\\
\textbf{Absorption} defined as the total engagement in the present experience \cite{brockmyer2009development}. In contrast to presence and in common with the flow, being in a state of psychological absorption induces an altered state of consciousness \cite{ ryan2000self}. In this altered state, there is a separation of thoughts, feelings, experiences and effect are less accessible to consciousness \cite{glicksohn1997explorations, sinclair2010testing}. While this feature is perceived as having a negative psychological effect, we are looking at its potential to detach and distract the players from feeling the discomfort associated with their impairment and, thus, avoid frustration and giving up.
An Engagement Questionnaire (EQ) was developed to evaluate user involvement in the ReHabgame using a self-report measure. The questionnaire has two parts, in part 1, a set of 16 questions require the user to rate on a Likert scale their experience in terms of flow (questions~3,~6,~8,~9,~16), presence (questions~5,~12,~13,~14,~15) and absorption (questions~1,~2,~4,~7,~10,~11). In the second part of the questionnaire, open-ended questions were asked, so that the players could explain their experience and provide feedback. The 5-point numeric rating scale is defined in that; 4 is "always agree", three is "sometimes agree", two is "neutral", one is "sometimes disagree", and 0 is "always disagree". The questions were listed in order from less engaged to more engaged ones. The answers provided for part 1 of the questionnaire were used to perform a psychometric analysis to determine the validity of the measurements. The part 2 of the questionnaire will be analysed and used to develop a beta version of the game in the future work.\\
Part one of the Engagement Questionnaire (EQ):
\begin{enumerate}
\item I feel spaced out (Absorption)
\item I feel intimidated (Absorption)
\item I get wound up (Flow)
\item The game kept me on my toes but did not overwhelm me (Absorption)
\item My actions seem to happen automatically (Presence)
\item I play without thinking about how to play (Flow)
\item I get into the game (Absorption)
\item I am so involved in what I am doing and cannot see myself separate from what I am doing (Flow)
\item If someone talks to me, I don’t hear them (Flow)
\item I lose track of time (Absorption)
\item I feel different (Absorption)
\item When moving through the game world, I feel as if I am there (presence)
\item I experience feelings as deeply in the game as I have in real life (Presence)
\item When playing the game, I feel as if I am an important participant (Presence)
\item I felt competent and effective (Presence)
\item Playing makes me feel calm (Flow)
\end{enumerate}
EQ-part two, answer the questions below in your terms: 
\begin{enumerate}           
\item What do you think about the appropriateness of the
length of time the rehabilitation took?
\item During the rehabilitation session, did you focus on the remaining time?
\item How exhausted do you feel?
\item How difficult did you feel today's session was?
\item How interesting did you find the game?
\item How quick did the time pass during the session?
\item How difficult was it to focus on the game?
\item How difficult was it to play the game?
\item Do you ever become so involved in the game as if you are inside the game rather than moving your limbs?
\item Please, tell us how we can improve the therapy sessions or the ReHabgame;
\end{enumerate}
\subsection{Subjects}
The pilot testing was conducted with $20$ post-stroke individuals (8 women, 12 men; mean age 54; range 41$-$63 years) who were recruited voluntarily. All participants were discharged from a hospital, $5.0 \mp 1.5$ years ago. The ethical approval was obtained before starting the study, from the University Computing and Technology Department’s Research Ethics committee. 
First, we described the study procedure and pattern of engagement and invited participants to read and sign an informed consent form. The study was continued for eight consecutive weeks in that player's joint velocity, acceleration, a ROM, head and spine tilt and hand gestures were collected and saved on the hard disk. The EQ was designed and administered post-play to determine the player's engagement in the game. \\

\subsection{Procedure}
The upper limbs (arm/hand) are designed to perform very skilled movements which allow us to do day to day activities. According to the therapist, to allow us to perform all of the intricate activities, the brain is wired to the upper limb for speed and accuracy \cite{ weinstein2018understanding}. This process makes it vulnerable to an injured person, and consequently, rehabilitation can be slow \cite{ramos2013brain}. The rehabilitation system has adapted an algorithm to allow the patients to use their arm and hand in ways that they would otherwise be unable to do so. The games in this study are the computer-generated simulation of 3D images and environments that can be interacted with in a seemingly real and physical way by a player. It does not use any head-mounted equipment such as Oculus/Google to avoid any motion sickness or discomfort for players (patients). The player's action partially determines what happens in the environment. The games were developed with direct collaboration and consultancy with physiotherapists and literature \cite{pedraza2015rehabilitation, hung2016stroke, fluet2010interfacing, ortiz2014virtual, kim2005swot}. The consultancy was performed throughout the development process as well as before starting to build the game. The information was gathered by asking the therapists what is important to be considered when we design our game and what elements must be implemented? Why don’t people attend their physiotherapy sessions? Why do they find it boring? etc.
We use the virtual rehabilitation system to promote rehabilitation and maximise the outcomes of the therapy. It aims to keep the muscles and joints of the hand and arm as healthy as possible, and this remains a key focus throughout the rehabilitation. This physiotherapy seeks to achieve as much function as possible and work towards an individual’s own goals. 

\begin{figure*}[thpb]
      \centering
     \includegraphics[scale=1.3]{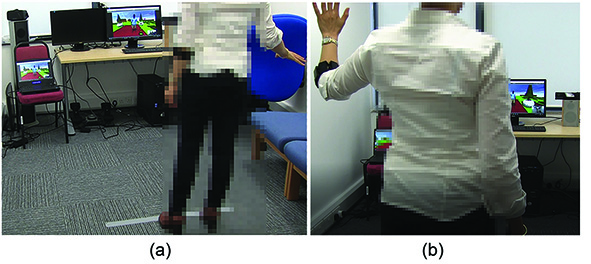}
      \caption{(a) Illustrates the screen shot of the video footage taken from the player while standing behind the predefined line within $1.8~m$ from the screen and playing the fruit “Reach-Grasp-Collect” game, (b) shows the same player with the Myo armband in her arm and the video streamed from behind the player.}
      \label{Fig.Player}
   \end{figure*}
Three/four minutes before starting the sessions the Myo armband was fitted to the participant's arm to warm-up and synchronise with the game and computer. Patients then were asked to stand within 1.5$-$2.0~m from the screen facing the Kinect as illustrated in Fig~\ref{Fig.Player}. When the player stands in the Kinect's field-of-view, the avatar figures are automatically generated. The participants were informed that the task is to interact with virtual objects generated on the screen. The virtual object's trajectories were stored on the hard disk which enables the estimation of joint centres using IK. The postural control or balance was monitored while interacting with the virtual objects through the avatar and player acts as a controller. The reach tests were conducted with identical instructions, with the subject advised that the movement of the player won't help in reaching object (the avatar's spine is locked in a global coordinate system). The players must keep their body and shoulders straight, show palm to the screen and close fingers for collection. Both games continued for one hour under the supervision of a medical scientist and a psychologist to assess the usability of the games for the target group.\\
\section{Results and Data Analysis}
The ROM and physical parameters, such as joint positions, orientations and relative angles were captured by the Kinect. The Kinect was used due to its possibility to enable an efficient and continuing rehabilitation programme in the home environment. It makes treatment more enjoyable thus increasing motivation and therefore adherence. The system has been developed so that the physical presence of a therapist is not required during the session and there is no need for the player to wear any marker or wires to be connected to sensors. The inertial measurement unit (IMU) and muscle activity were collected via the Myo with the blue tooth connection. The data were used by the algorithm and also saved for off-line analysis. The player receives real-time feedback on performance as the MCTS simulation incorporates data efficiently from the player's previous performances for a subsequent generation in the game. The motion data from the Kinect and Myo undergo a process of smoothing to achieve higher accuracy and efficiency. A smoothing algorithm was used to remove any noise from the signals collected by the sensors. The angular measurements and accurate linear motions were obtained by the IMU (gyroscope, accelerometer and magnetometer) sensors to distinguish precise linear movement from the accelerometer readings. 
The EQ and ROM data were collected from each of the study participants. The participants played the ReHabgame with small breaks between each one to prevent fatigue. All feedback, suggestions and observations were collected to improve the game and tested again by the same persons and using the same EQ. The data gathered from the various stages of the therapy sessions were used for qualitative analysis. \\
Rasch analysis was implemented using Ministep $4.0.1$ (Winsteps) software\footnote{http://www.winsteps.com/ministep.htm}. In 1969, Georg Rasch\footnote{https://www.rasch.org/} proposed a statistical model that complied with the fundamental assumptions made in measurements in neuroscience and has been used to aid in the construction and validation of health status questionnaire in neurology
\cite{rasch1993probabilistic, tesio2003measuring, duncan2003rasch}.\\
\begin{table}
\centering
\setlength{\tabcolsep}{0.8pc}
\caption{The Rasch analysis of Engagement Questionnaire (EQ) results (part one).}
\label{Fig.table1}
\begin{tabular}{|c|c|c|c|c}
\hline
\multicolumn{4}{c}{Results of the Rasch rating scale model (RSM)}\\
\centering
Item       & ~~Item difficulty &~ Item Fit (mean squares)&RMSR (root-mean-square residual) \\
         \cline{3-3}
       
No.  & ~~~~ (Logit) &Infit ~~~~~~~~~~~~~Outfit ~~~&~~~~~~~(observation - expectation)\\
\hline
(3)       & ~~~~~ 1.13     & 1,48 ~~~~~~~ ~~~~~~~1.52&~~~~~~1.16 \\
(6)       & ~~~~~ 0.98     & 1.47 ~~~~~~~ ~~~~~~~1.44&~~~~~~1.20\\
(12)       & ~~~~~-0.37     & 1.31 ~~~~~~~ ~~~~~~~1.28&~~~~~~1.05 \\
(15)       & ~~~~~-0.85     & 1.16 ~~~~~~~ ~~~~~~~1.23&~~~~~~0.85 \\
(4)       & ~~~~~-0.05     & 1.15 ~~~~~~~ ~~~~~~~1.11&~~~~~~1.08 \\
(9)       & ~~~~~-0.05     & 1.08 ~~~~~~~ ~~~~~~~1.06&~~~~~~1.04 \\
(8)       & ~~~~~-0.31     & 0.98 ~~~~~~~ ~~~~~~~1.04&~~~~~~0.93 \\
(16)       &~~~~~~-1.02     & 1.00 ~~~~~~~ ~~~~~~~0.94&~~~~~~0.75 \\
(5)       & ~~~~~ 0.88     & 0.99 ~~~~~~~ ~~~~~~~0.92&~~~~~~1.01 \\
(1)       & ~~~~~ 1.25     & 0.97 ~~~~~~~ ~~~~~~~0.90&~~~~~~0.91 \\
(7)       & ~~~~~-0.37     & 0.95 ~~~~~~~ ~~~~~~~0.93&~~~~~~0.90 \\
(2)       & ~~~~~ 0.64     & 0.88 ~~~~~~~ ~~~~~~~0.85&~~~~~~0.99 \\
(11)       & ~~~~~-1.02     & 0.88 ~~~~~~~ ~~~~~~~0.86&~~~~~~0.70 \\
(14)       & ~~~~~-0.43     & 0.81 ~~~~~~~ ~~~~~~~0.80&~~~~~~0.81 \\
(13)       & ~~~~~-0.31     & 0.79 ~~~~~~~ ~~~~~~~0.74&~~~~~~0.83 \\
(10)       & ~~~~~-0.10     & 0.68 ~~~~~~~ ~~~~~~~0.68&~~~~~~0.82 \\
\hline
\multicolumn{4}{@{}p{110mm}}{Item Mean (SD): 45.8 (14.1), and Person Mean (SD): 36.6 (7.3)}
\end{tabular}
\end{table}
In EQ, 16 multiple choice questions were developed with 5-point numeric rating scale answers from $0$ (always disagree) to $4$ (always agree). Due to the items’ equal rating scale structure, the Rash Rating Scale Model (RSM)~\cite{andrich1978rating} was used to determine the participants’ engagement in the ReHabgame \footnote{https://www.rasch.org/rmt/rmt94n.htm}. Unidimensionality, reliability, targeting and response categories were examined.\\
To assess the unidimensionality (i.e. fitness for the Rasch model), a principal component analysis of Rasch residuals was used in which the observed response is compared with the predicted response of the model. Under good fit conditions for rating surveys, infit and outfit mean squares must be within the range of $(0.6$ to $1.5)$ \cite{millis2014rasch}. The internal consistency of person and item performance were studied using separation reliability estimates and separation ratios. Separation reliability for person refers to the consistency of a person$'$s responses across items. The item$'$s separation reliability refers to the consistency of item performances across persons. Separation ratios $> (1.5$ to $2)$ provide evidence of internal consistency.
Targeting was done using an examination of the item/person map in which the distribution and spread of items and persons along an ordinary logit were compared. Targeting is measured as a tool for participants’ engagement and is defined as the extent to which items are of appropriate difficulty for the sample or item endurability to a person’s level of engagement. A logit is a Log-odds unit or a logarithmic transformation of the ratio of the probabilities of a correct and incorrect response scale of the latent construct or the probabilities of adjacent categories on a rating scale. Cronbachs’ alpha (Rasch Person Reliability) was used to measure the internal consistency between items and its strength \cite{duncan2003rasch}. Item Response Theory (IRT) model or the probability of an individuals’ response to an item and model item characteristic curve (ICC) or the relationship between the probability of success to an item were assessed. \\

\subsection{Results}
\begin{figure*}[thpb]
      \centering
     \includegraphics[scale=0.4
     ]{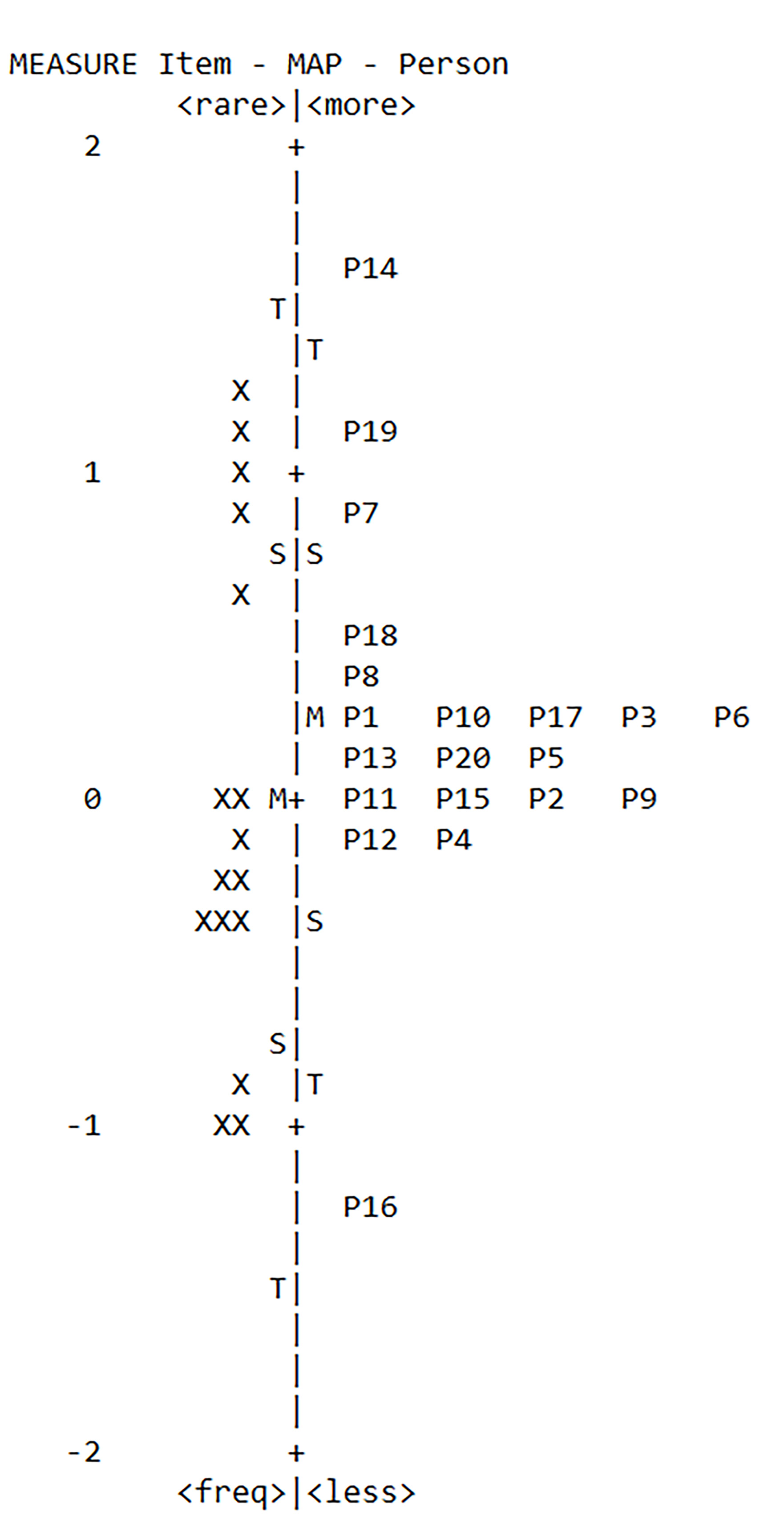}
      \caption{Participants’ engagement scores (each x represents one person), centred at a mean of zero with standard deviations of $1.0$. Items are arranged in engagement order with $4$ referring to "always agree" and $0$ to "always disagree".}
      \label{Fig.rasch_table}
   \end{figure*}

\begin{figure*}[thpb]
      \centering
     \includegraphics[scale=0.6
     ]{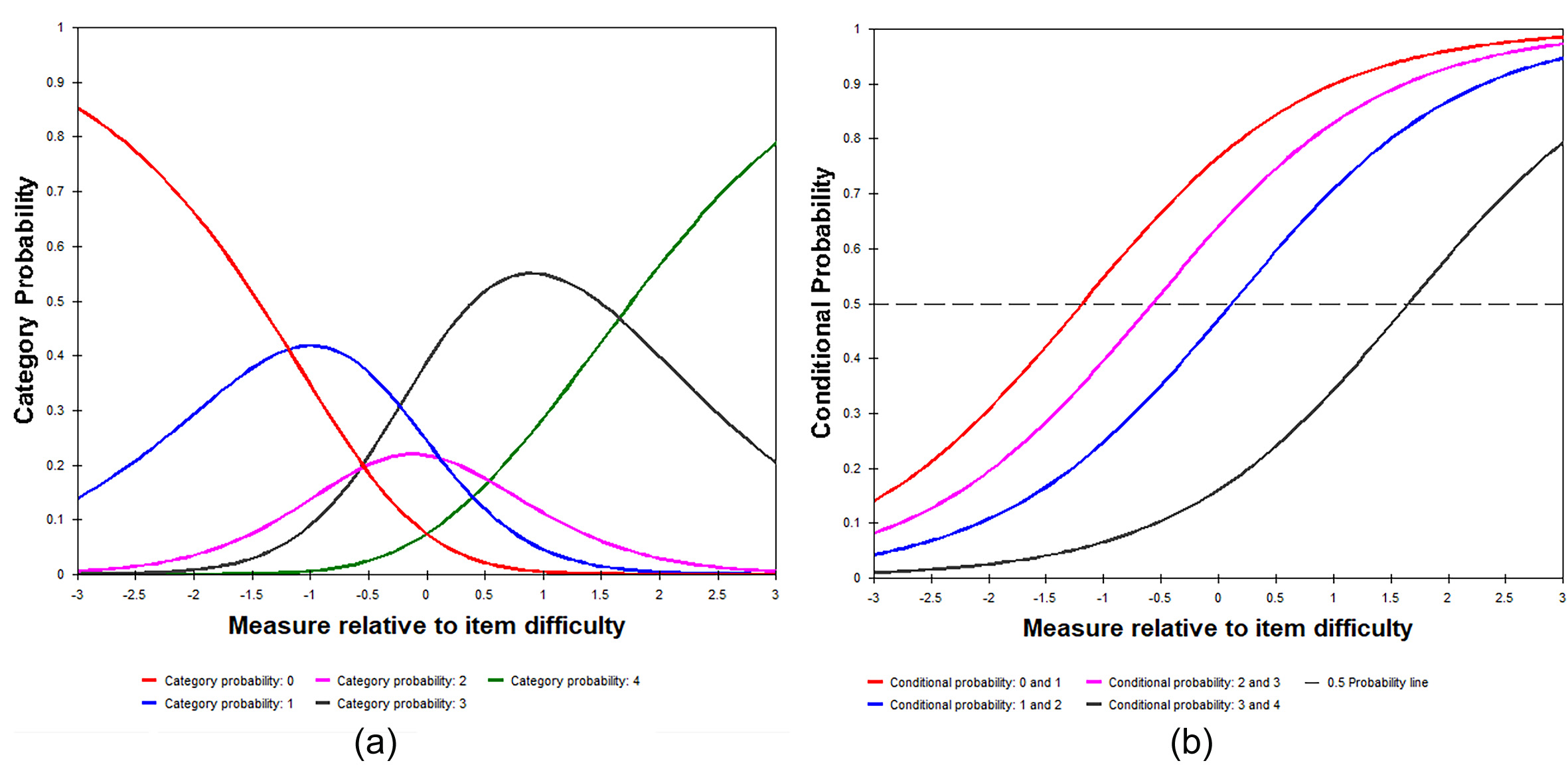}
      \caption{The category probability curves for items test information for engagement in all categories (a) and their probabilities (b).}
      \label{Fig.rasch}
   \end{figure*}

\begin{figure*}[thpb]
      \centering
     \includegraphics[scale=0.6
     ]{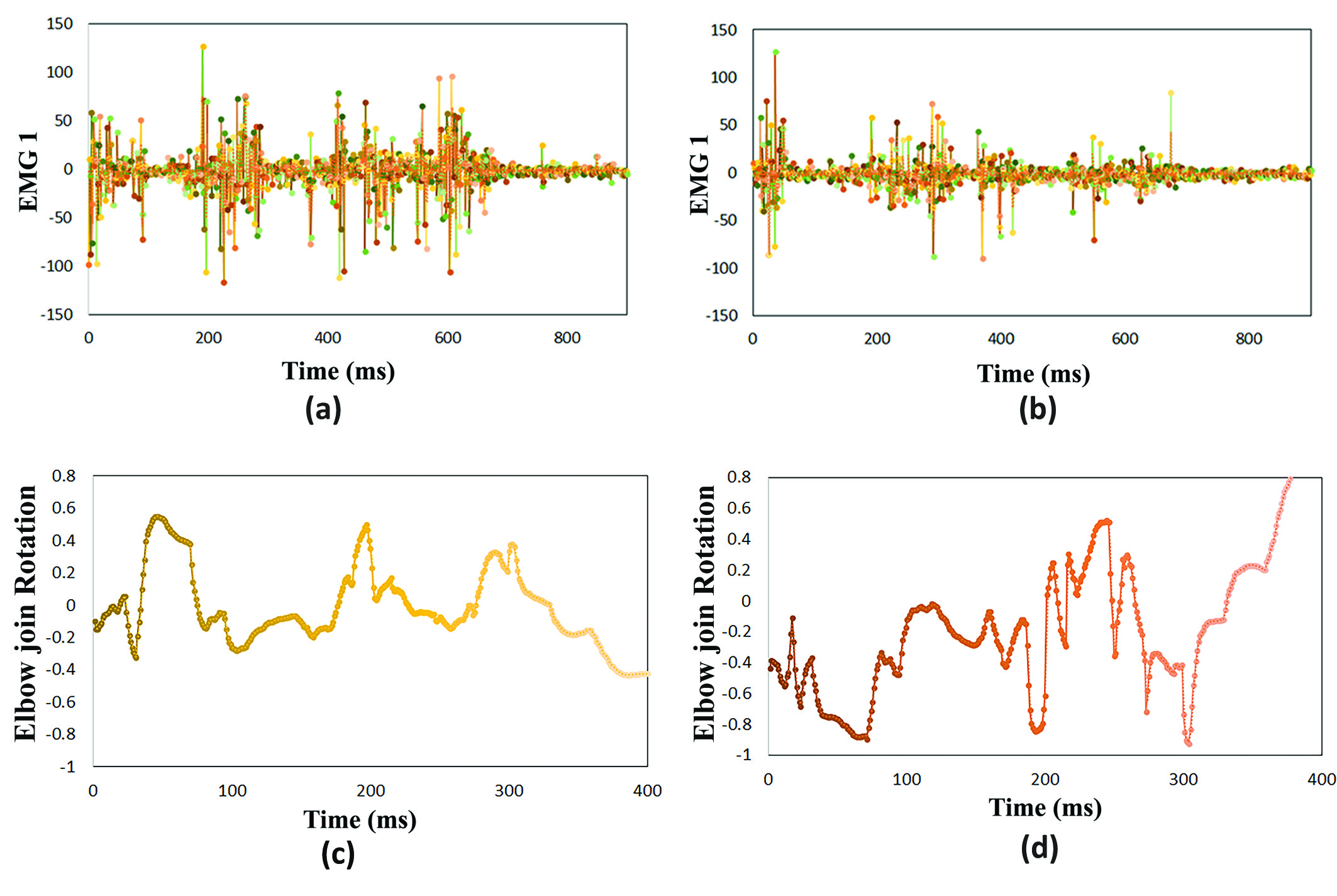}
      \caption{(a) and (b) show data collected from Myo armband of the electromyography (EMG 1) generated through “Reach-Press-Hold” buttons game"  from two players, (c) and (d) illustrates the elbow joint orientation collected by the Kinect from the same players.}
      \label{Fig.EMG}
   \end{figure*}

The Rasch analysis and RSM are outlined in Table~\ref{Fig.table1}. It shows the item numbers which correspond to the question numbers, the items’ difficulty (Logit), infit and outfit mean squares and RMSR for each item. Infit and outfit mean squares values are in the range of (0.68$-$1.52) for all 16 items, which satisfy the unidimensionality condition. The initial fit analysis statistics for the individual items show that the easiest items were "(11): I feel different (Absorption)", and "(16): Playing makes me feel calm (Flow)" with the logits equal to $-1.02$. The most difficult item was item number "(1): I feel spaced out" with the highest logit value $1.25$. RMSR has its highest value of $1.2$ for item "(6)" and the lowest value of $0.7$ for item "(11)". The Ministeps item/person map is shown in Fig~\ref{Fig.rasch_table}. Item/person map would have a good targeting when the items and persons are clustered symmetrically all around the vertical axis. The item measures ranged from $-1.02$ to $1.25$ logit and persons’ logit was in the range of $-1.25$ to $1.64$. Therefore, the Item/person map has a good targeting of items. The left side of the Fig~\ref{Fig.rasch_table} represented by $"X"$ illustrates the items and the right side are the participants ($P1$-$P20$) with good distribution to characterise the participants’ engagement.\\
The Rasch expectation is that individuals with higher engagement in a game would have a higher likelihood of agreeing with more engaged items in the categories (higher response category). Fig~\ref{Fig.rasch} sketched out the category and conditional probability curves relative to item difficulty. The peaks of the curves for the $5$ categories ("always disagree = 0" to "always agree = 4") are always in ascending order along the latent variable. The crossover point for the category $2$ is within category $1$ and $3$, which means this category could be collapsed into category $1$ and $3$. The person/item threshold distribution map is outlined in Fig~\ref{Fig.histogram} with the x-axes displaying location or difficulty of item thresholds and the y-axes display the frequencies of item thresholds and participants. It can be seen that the offset of person to item suggests targeted scale.\\
Some of the important findings taken from the second part of the EQ are listed below. The majority of participants enjoyed the rehabilitation sessions and the ReHabgame activities. Overall, the participants commented very positively and reported that the activities covered a good ROM for the upper body. They suggested that audio feedback could be useful to create a more exciting environment in the game.\\
Fig \ref{Fig.EMG}a, Fig \ref{Fig.EMG}b illustrate the muscle activity during the gameplay collected by one of the EMG sensors of Myo from two participants while playing the same game. Fig \ref{Fig.EMG}c, Fig \ref{Fig.EMG}d shows the elbow rotation data generated in the hard disk and collected from the Kinect.
   
\begin{figure}
\centering
    \begin{subfigure}{0.4\textwidth}
        \centering
        \includegraphics[width=\linewidth]{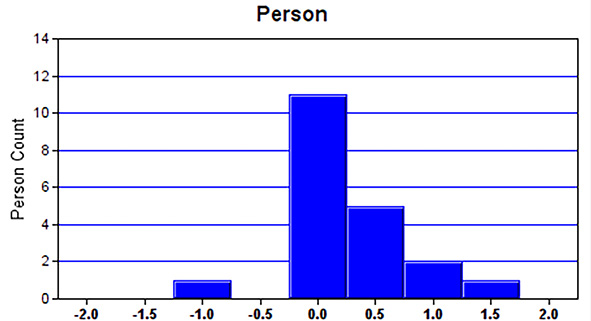}
        \caption{}\label{Fig.histogram}
    \end{subfigure} %
    \begin{subfigure}{.4\textwidth}
        \centering
        \includegraphics[width=\linewidth]{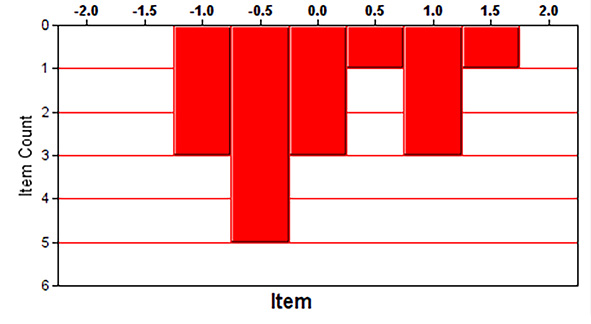}
        \caption{}\label{Fig.histogram}
    \end{subfigure} %
       \caption{The threshold map with the distribution of person in figure (a) and item in figure (b).}
\end{figure}
Some of the feedback \& suggestions received from players are listed below:
\begin{enumerate}
\item \boldmath{"I like to play longer, and I feel it helps me to score better"}
\item \boldmath{"The game becomes intense occasionally but still motivating, and I like to improve my score each time"}
\item \boldmath{"The ReHabgame makes my muscles to be used [and] without realising they are getting stronger"}
\item \boldmath{"The positive impact of the game was noticed by third parties (family members and my physiotherapist)"}
\item \boldmath{"I found it very interesting. I feel the positive effects of the exercise on my whole body"}
\item \boldmath{"My whole body was engaged with the game, and I feel I have more control over my posture"}
\item \boldmath{"The difficulty level of the game seemed to match my ability, and I score better each session"}
\item \boldmath{"The more I play I find that my finger movements are improving"}
\item \boldmath{"I like that the gameplay was challenging but not exhaustive"}
\item \boldmath{"I needed to concentrate a lot, and I couldn't hear what people were saying around me"}
\item \boldmath{"The game helps and improves the upper limb reaction and reduces its rigidity."}
\item \boldmath{"I found the visual feedback very instructive, but adding sound to the game would make it more stimulating"}
\item \boldmath{"I would like to be able to compare my performance and score with previous sessions on graphs"}
\item \boldmath{"I prefer VR games like the ReHabgame since head-mounted displays in other VR systems cause motion sickness"}
\end{enumerate}
\section{Discussion}
\subsection{Summary of main findings}
To our knowledge, our study is the first to develop the rehabilitation game with the capabilities build in the ReHabgame. A Rasch model was developed to evaluate the participant engagement in the game as well as the system effective and attractiveness. 
The developed system enables a therapist to present a variety of controlled stimuli to measure and monitor the progress of the participants. The finding showed that the therapy provides an adequate physical and mental engagement through sufficient visual feedback throughout the sessions. It gives the opportunity for patients to perform additional exercises independently. It offers the possibility for a clinician to recommend an appropriate and individually aligned exercise. The attractiveness of the ReHabgame was measured by the standard flow, presence and absorption (psychological model balancing the player’s perceived skill with the perceived challenge). The effectiveness was assessed to evaluate the physical balance between fitness (the body’s ability in tolerating exercise), and intensity (the difficulty of the training on the body). The Rasch analysis in this study showed that a unidimensional construct was created and fit statistics were obtained for the measured properties.
The ReHabgame successfully calibrated the ROM of the players, muscle activities, linear and angular motions. The required movement or exercise level was adopted through the MCTS and the ROG based on HSS and MAS. In that skill level and game difficulty were gradually increased to challenge the player.
\subsection{Strengths and limitations}
Our study had some limitations in that the sample of participants were not fully representative of the diverse population of stroke patients (41 $-$ 63) years old which was not wide enough to include younger or older people. The percentage of female participants to male was $0.4$. Additionally, the ethics approval process to have access to the NHS was cumbersome and time-consuming therefore we decided to limit our study to the volunteers of communities. 

\section{Conclusion}
The feasibility and effectiveness of the ReHabgame were studied as stimuli for improving the motor performance in a post-stroke participant. It facilitates autonomous physical rehabilitation at patient's state as prescribed by the therapist. This system provided an environment that encourages patients to build greater strength and endurance on tasks that increase independence and enrich daily activity. The system monitors the trajectory of the hand during movements about the shoulder, arm, and elbow. The data taken from the devices are saved and accessed for further analysis as well as to generate the best game suitable for a patient with a certain disability through the intelligent computing system. The algorithm generates objects randomly and monitors the execution of the exercise, and the system provides direct feedback in real-time accordingly.
The therapy provided an adequate physical and mental engagement that offered sufficient visual feedback throughout treatment sessions. The MCTS and ROG algorithms were used to control the intensity of the physical activity that happened gradually based on the player's performance. The data were obtained at devices local sampling frequency followed by performing linear interpolations to create a constant sampling interval while maintaining time and frequency domain signal integrity to achieve an identical sampling frequency to the kinematic data.\\
The subjective experience was conducted through the EQ to gather further information from the subjects about their experience of game playing. Some positive elements of the ReHabgame that was claimed by the player were the beautiful scenery and appealing feedbacks they received which had a direct relation to the engagement of a player.
It aimed to distract the player's attention towards the exciting aspects of the game to avoid fatigue. The training engages users to allow for the constant intensive practice required for motor plasticity. It monitors and progressively corrects the abnormalities of the kinematic movement, coordination and assess the transmitted electrical signals as the muscles strain. The subjective experience was performed through the EQ to gather further information from the subjects about their experience in playing the games. Future work includes experiments involving individuals with motor impairment, with a specific focus on gait and balance. It also aims to incorporate the possibility of implementing social interactions through online gameplay at patient's home or could be arranged in social communities by patients carer or charity organisation. That increases the motivation by gathering more patients to cooperatively (or competitively) interact and play the game. It could enable a long-term recovery and could lead to better rehabilitation outcome \cite{hung2016stroke, pedraza2015rehabilitation}. The open-ended questionnaire also will be used for developing the beta version of the ReHabgame in the future work.

\section*{Acknowledgement}
The authors acknowledge Anglia Ruskin University for funding, organising and hosting the pilot study and testing interventions.
\section*{Reference}
\bibliographystyle{elsarticle-harv}
\bibliography{Ref}
\end{document}